 \documentstyle[aps,prl,twocolumn]{revtex}

 \draft

\begin{document}
\title{Thermal  vs quantum decoherence in double well trapped Bose-Einstein
condensates}
\author{L. Pitaevskii$^{a,b}$ and S. Stringari$^{a}$}
\address{$^{a}$Dipartimento di Fisica, Universit\`{a} di Trento,}
\address{and Istituto Nazionale per la Fisica della Materia,}
\address{I-38050 Povo, Italy}
\address{$^{b}$ Kapitza Institute for Physical Problems,} 
\address{ ul. Kosygina 2, 117334 Moscow, Russia}
\date{\today}
\maketitle

\begin{abstract}
The quantum and thermal fluctuations of the phase are investigated in a cold
Bose gas confined by a double well trap. The coherence of the system is
discussed in terms of the visibility of interference fringes in both
momentum and coordinate space. The visibility is calculated
at zero as well as at finite temperature. 
The thermal fluctuations are shown to affect
significantly the  transition from the coherent to the incoherent
regime even at very low temperatures. The coherence of an array of multiple
condensates is also discussed.
\end{abstract}

\bigskip

\narrowtext

After the first interference measurements carried out on two expanding and
overlapping condensates \cite{mitinterf} the problem of the relative phase
between Bose-Einstein condensates has stimulated extensive theoretical work
in the last few years (see for example \cite{leggett} and references
therein) as well as new experimental attempts to point out Josephson like
phenomena  \cite{jila,kas}. A
challenging question is the transition from a coherent to an incoherent
regime associated with the increase of the fluctuations of the relative
phase between condensates confined in different traps. The transition can be
in principle controlled by changing the tunneling probability between the
wells either by tuning the height and the width of the barrier or the number
of atoms. Most of theoretical works have focused so far on the ideal case of
zero temperature where the fluctuations of the phase are of quantum nature.
These works have shown that if the tunneling probability is sufficiently
small the ground state of an atomic gas interacting with repulsive forces is
not globally coherent, but exhibits new features of quantum
nature, associated with the appearance of number squeezed configurations
(see for example \cite{java}).

The purpose of this letter is to discuss the relative importance of the
quantum and thermal fluctuations  in driving the transition
between the coherent and incoherent regimes. Due to the low energy scale
associated with the Josephson oscillation of the condensates the thermal
effect may actually become crucially important even if one works at very low
temperature.

Let us consider a dilute gas of atoms confined by an external
potential characterized by a double well 
along the $x$ direction. The problem of an array of multiple wells
will be discussed in the last part of the work. Near the bottom of the two
traps the potential can be approximated by a harmonic function characterized
by the oscillator frequency $\omega_{ho}$ which provides the relevant
frequency scale of the collective excitations of each condensate
(actually, due to the 3D nature of the problem, the harmonic
expansion will in general introduce three different frequencies). We start
our discussion by  recalling the classical problem of the Josephson
oscillation in the framework of the Gross-Pitaevskii (GP) theory for the
order parameter. If the overlap between the condensates confined in the two
wells is small we can naturally construct an approximate solution of the
time dependent GP equation in the form 
\begin{equation}
\Psi \left( {\bf r,}t\right) = \left[ \Psi _{l}\left( {\bf r},N_{l}\right)
e^{i\Phi_{l}}+\Psi _{r}\left( {\bf r},N_{r}\right) e^{i\Phi_{r}}\right] \, ,
\label{PsiJ}
\end{equation}
where $\Psi _{l,r}\left( {\bf r},N_{l,r}\right)$ are real functions
satisfying the stationary GP equation in the left ($l$) and right ($r$)
wells respectively and normalized to $\int d{\bf r} \Psi_{l,r}^2 = N_{l,r}$
with $N_l+N_r=N$. The time dependence of the solution (\ref{PsiJ}) is
contained in the phases $\Phi_{l,r}$ and in the number of atoms $N_{l,r}$
confined in each well. The equations of motion can be written in the
canonical form $\partial \Phi /\partial t=\partial H_{J}/ \partial (\hbar k) 
$ and $\partial( \hbar k) /\partial t=-\partial H_{J}/\partial \Phi$ where 
\begin{equation}
H_{J}={\frac{1}{2}}E_{C}k^{2}-E_{J}\cos \Phi \, .  \label{JosHam}
\end{equation}
is the so called Josephson Hamiltonian \cite{smerzi,zapata} depending on the
conjugate variables $\hbar k$ and $\Phi$ where $k=\left( N_{l}-N_{r}\right)
/2$ is determined by the relative number of atoms in the two condensates and 
$\Phi = \Phi_l-\Phi_r$ is their relative phase. In deriving eq.(\ref{JosHam}%
) we have assumed $k \ll N$ and introduced the relevant parameters $E_{c}=2
d\mu _{l}/dN_{l}$ and $E_J = (\hbar^2 / 2m) \int dy dz [\Psi_l\partial
\Psi_r/ \partial x - \Psi_r \partial \Psi_l / \partial x]_{x=0}$ where $%
\mu_l $ is the chemical potential of each condensate and all the quantities
are evaluated at $N_l=N_r=N/2$. The term in $E_c$ accounts for the
interaction effects in each condensate and vanishes in the non
interacting gas since in this case the chemical potential does not depend on
the number of atoms. In the Thomas-Fermi limit instead the parameter $E_C$
takes the value $E_C=(4/ 5)\mu_l /N_l$ due to the $N^{2/5}$ dependence of
the chemical potential of a Bose-Einstein condensate trapped by a 3D
harmonic potential. In the Thomas-Fermi regime the value of $E_C$ hence
decreases with $N$. The term in $E_J$, which describes the tunneling
probability between the two wells and is at the origin of the Josephson
current, instead increases with $N$. Expression (\ref{JosHam}) for $H_J$
 can be easily 
generalized to include the effects of  gravity.

Equation (\ref{PsiJ}) represents a good approximation to the solution of the
GP equation only if the Josephson motion is decoupled from the other modes
of the condensate. This implies that the quantities $\Phi$ and $k$ should
vary slowly in time with respect to the typical time $1/\omega_{ho}$
characterizing the internal motion of each condensate. The Hamiltonian (\ref
{JosHam}) corresponds to the classical problem of the pendulum. For small
oscillations near equilibrium ($k=0$, $\Phi=0$) the motion is of harmonic
nature and is characterized by the classical plasma frequency 
\begin{equation}
\omega_{cl} = {\frac{1}{\hbar}}\sqrt{E_JE_C} \, ,  \label{omegap}
\end{equation}
so that, in order to ensure the decoupling from the internal oscillations
of the condensate, the inequality $\omega_{cl} \ll \omega_{ho} $ should be satisfied.

Let us now proceed to quantize the classical Hamiltonian (\ref{JosHam}).
This is achieved by replacing the conjugated variables $\Phi $ and $\hbar k $
with operators satisfying the commutation relation $[\hat{\Phi},\hbar \hat{k}%
] = i\hbar$. It is convenient to work in the ''$\Phi $-representation''
where $\hat{k}=-i\partial /\partial \Phi $ and the quantum problem can
be described by the Hamiltonian 
\begin{equation}
\hat{H}_{J}=-\frac{E_{C}}{2}\frac{\partial ^{2}}{\partial \Phi ^{2}}%
-E_{J}\cos \Phi  \label{HJquant}
\end{equation}
acting in the space of the periodical functions of period $2\pi$. The
quantization introduces quantum fluctuations in the equilibrium state of the
system whose nature depends on the ratio $E_C/E_J$. Because of the
periodicity constraint the uncertainty relation obeyed by the fluctuations
of $\Phi$ ad $k$ can be conveniently written in the form 
\begin{equation}
\langle \Delta k^2\rangle \langle \Delta \sin^2\Phi \rangle \ge
\langle\cos\Phi\rangle ^2 /4  
\label{URphi}
\end{equation}
where $\langle \Delta A^2\rangle = \langle \hat{A}^2 \rangle - \langle \hat{A%
} \rangle^2$ is the variance of the general observable $\hat{A}$. The
uncertainty relation (\ref{URphi}) reduces to the usual form $\langle\Delta
k^2\rangle \langle\Delta \Phi^2\rangle \ge 1/4$ only if  the
phase is localized around $\Phi= 0$ \cite{note1}. The quantity 
\begin{equation}
\alpha = \langle\cos\Phi\rangle  \label{alpha}
\end{equation}
characterizing the right hand side of the uncertainty inequality (\ref{URphi}%
) has a physical meaning of first importance. It provides the degree of
coherence of the configuration and will be called the {\it coherence}
factor. If the value of the phase is localized around zero, the value of $%
\alpha$ is close to unity (full coherence) and one can expand eq.(\ref{alpha}%
) to calculate the quadratic fluctuations of the phase as $\langle\Delta \Phi^2\rangle
 = 2\left( 1-\alpha \right)$. If instead the phase is delocalized and
all its values are equally probable, then the value of $\alpha$ is zero
(absence of coherence).

The parameter $\alpha$ is directly related to measurable observables. For
example if one calculates the {\it in situ} momentum distribution of two
separated condensates with a fixed relative phase $\Phi$, one finds fringes
in momentum space given by the expression \cite{PS}  $n\left( {\bf p}\right)
= 2\left[ 1+ \cos \left( p_{x}d/\hbar + \Phi\right)\right] n_0({\bf p})$ 
where $%
n_0({\bf p})$ is the momentum distribution of each condensate and $d$ is the
relative distance along the $x$ axis. The typical width of $n_0({\bf p})$ is
given by $\hbar/R$ where $R$ is the spatial size of each condensate. The
momentum distribution of a single trapped Bose gas has been already the
object of measurements based on stimulated light scattering \cite{mitnp}. If
the value of the relative phase fluctuates the average value of the momentum
distribution at equilibrium is described by the equation 
\begin{equation}
n\left( {\bf p})\right) = 2\left[ 1+\alpha \cos \left( p_{x}d/\hbar \right)\right]
n_l({\bf p})  \label{phint-av}
\end{equation}
where the parameter $\alpha$ corresponds to the average $\langle\cos\Phi%
\rangle$ and hence coincides with the coherence factor introduced in eq.(\ref
{alpha}) (we have used the fact that $\langle\sin\Phi\rangle=0$ at
equilibrium). One should not however confuse the parameter $\alpha$ with the
visibility of fringes obtained in a {\it single} realization of the
experiment which can give rise to a well defined interference pattern even
if the initial state is not coherent. The parameter $\alpha$ corresponds
instead to an {\it average} taken on several measurements and is
consequently an intrinsic property of the state of the system. This
distinction is crucial if one considers interference experiments with two
condensates. A similar structure characterizes also the fringes of the
density distribution measured after expanding the condensates \cite
{mitinterf}. In this case in order to obtain a simple 
result one has to make the additional
assumption that the condensates do not interact each other when they
overlap. For long expansion times $t$ the phases of the two condensates 
evolve according to the law  $({\bf r} \pm {\bf d}/2)^2m/\hbar t$ \cite{RMP}
so that the full  density profile takes the following asymptotic form
\begin{equation}
n\left( {\bf r},t)\right) = \left[n_+ + n_-
+ 2\alpha \cos \left({\frac{md}{\hbar t}}
x \right)\sqrt{n_+n_-}\right] 
\label{intnr}
\end{equation}
where $n_{\pm}=n_0({\bf r} \pm {\bf d}/2,t)$ are the densities of the two 
expanding condensates. Eqs.(\ref{phint-av}-\ref{intnr}) show that the
value of $\alpha$ can be inferred from either the measurement of the
momentum distribution and/or of the density profile after expansion,
although in the second case the conditions of applicability of eq.(\ref
{intnr}) are more severe.

Let us now discuss the behaviour of the ground state of the Josephson
Hamiltonian (\ref{HJquant}). In the limit of {\it strong} tunneling $%
E_{C}/E_{J} \ll 1$ the system undergoes small oscillations around the
equilibrium value $\Phi =0$ and the Hamiltonian becomes quadratic in $\Phi$. 
In this limit the fluctuations of the phase are given by $%
\langle\Delta\Phi ^{2}\rangle = (\sqrt{E_{C}/E_{J}})/2 \ll 1$ and are hence small.
One can consequently regard the inequality $E_C/E_J \ll 1$ as the physical
condition for applying the classical GP theory to the problem
of the double-well potential and, consequently, to treat the systems as a
globally coherent object described by a unique order parameter. Also the
fluctuations of $k$ in the ground state can be easily calculated and one
finds the result $\langle \Delta k^{2}\rangle = (\sqrt{E_J/E_{C}})/2 \gg 1$. In the
strong tunneling limit the uncertainty relation (\ref{URphi}) reduces to the
usual form $\langle \Delta k^2\rangle\langle \Delta \Phi^2\rangle
 \ge 1/4$ and actually becomes an
identity as a consequence of the harmonic nature of the Hamiltonian. Notice,
however, that if $\langle\Delta\Phi^2\rangle$ is of the order of $1/N$ or
smaller, the formalism developed above is no longer adequate and should be
improved through both the inclusion of the factor $\sqrt{1-4k^2/N^2}$ in the
term $-E_J\cos\Phi$ of the Josephson Hamiltonian (\ref{HJquant}), and a more
microscopic approach to the phase operator (see for example \cite{ADS} and
references therein). In the following we will mainly limit our discussion to
the case $E_C\gg E_J/N^2$ which ensures that the quadratic fluctuations of
the phase are in any case larger than $1/N$.

In the case of {\it weak} tunneling $E_{C}/E_{J}\gg 1$ the behaviour of the
fluctuations is very different. The Josephson term $E_{J}$ entering the
Hamiltonian (\ref{HJquant}) can be neglected in first approximation and the
wave function of the ground state is simply a constant given by $1/\sqrt{%
2\pi }$, showing that the relative phases between the two condensates is
distributed in a random way. At the same time the fluctuation of the
relative number $k$ of atoms in the two traps becomes smaller and smaller
and vanish like $2 (E_J/E_C)^2$. This is consistent with the uncertainty
relation (\ref{URphi}) since in the same limit the coherence factor vanishes
like $\alpha \to 2(E_J/E_C)$. In the weak tunneling limit the ground state
of the system exhibits fragmented Bose-Einstein condensation and is
characterized by the macroscopic occupation of the two single-particle
states localized in the two wells.

In fig.1 we show the coherence factor $\alpha$ as a function of the ratio $%
E_C/E_J$ calculated by solving explicitly the Schr\"{o}dinger equation for
the ground state with the Josephson Hamiltonian (\ref{HJquant}). The figure
shows that for values of $E_J$ smaller than $E_C$ the coherence factor is
significantly quenched, pointing out the occurrence of a continuous
transition to the number-squeezed regime. The transition is accompanied by a change
of the lowest excitation energy $\omega_p$ of $\hat{H}_J$ from the 
classical value (\ref{omegap}) to the "free" value $E_C/2$ which is obtained by setting
$E_J=0$ in (\ref{HJquant}). An accurate expression for  $\omega_p$ is obtained
by evaluating the ratio between the cubic and energy weighted sum rules 
$\langle [[\hat{k},\hat{H}_J],[\hat{H}_J,[\hat{H}_J,\hat{k}]]]\rangle =E^2_JE_C\langle 
\cos \Phi^2 \rangle$ 
and $\langle [\hat{k},[\hat{H}_J,\hat{k}]]\rangle = E_J\langle 
\cos \Phi\rangle$,  relative
to the operator $\hat{k}$. The result is 
\begin{equation}
\omega_p = {1\over \hbar} \sqrt{E_JE_C {\langle \cos^2 \Phi\rangle \over \langle \cos \Phi\rangle}}
\label{omegapq}
\end{equation} 
where the averages should be evaluated on the ground state. In the limit $E_C \ll E_J$
one has $ \langle \cos ^2\Phi\rangle = \langle \cos \Phi\rangle =1$ and one recovers
the classical result (\ref{omegap}). In the opposite limit one finds 
$\langle 
\cos \Phi^2 \rangle = 1/2$, $\langle 
\cos \Phi \rangle = 2E_J/E_C$ and $\omega_p = E_C/2\hbar$.

So far we have investigated the problem in an ideal situation at zero
temperature. Due to the smallness of the plasma frequency (\ref{omegap}) it
is important to understand the role of the thermal fluctuations of the
relative phase of the two-condensate system. These are expected to become
important as soon as the temperature is of the order or higher than $\hbar
\omega_p$. To investigate the thermal effect we have calculated the
thermal average 
$\alpha(T) = \sum_n \alpha_n\exp[-E_n/T]/ \sum_n\exp[- E_n/T]$
of the coherence factor 
where $n$ and $E_n$ are the eigenstates and eigenenergies of the Josephson
Hamiltonian (\ref{HJquant}), $\alpha_n= \langle n\mid \cos\Phi\mid n\rangle$
are the corresponding quantum averages and we have set the Boltzmann
constant equal to $1$. The results are presented in fig. 2 as a function of
the parameter $T/E_J$ for two different values of $E_C/E_J$. At $T=0$ one
recovers the values of $\alpha$ given in fig.1. One clearly see that even if
quantum effects are small the thermal decoherence of the  phase
becomes important at temperatures of the order $T \sim E_J$. The full line
in the same figure gives  the classical prediction 
\begin{equation}
\alpha_{cl}(T) = {\frac{\int_{-\pi}^{+\pi} d\Phi \cos\Phi \exp[E_J\cos\Phi /T%
] }{\int_{-\pi}^{+\pi} d\Phi \exp[E_J\cos\Phi /T]}} 
  \label{alphacl}
\end{equation}
holding for $T \gg \omega_p$. If $T\ll E_J$, eq. (\ref{alphacl}) gives
the result $\langle\Delta \Phi^2 \rangle = T/E_J \ll 1$ for the classical
fluctuation of the phase. The fluctuation of $k$ can be also easily
calculated starting from the classical expression (\ref{JosHam}) and 
one gets the result $\langle\Delta k^2\rangle =
T /E_C$ for any value of $T$. The thermal
fluctuations of $k$ may be large and consequently at finite $T$ the 
system will not in general exhibit the phenomenon of number squeezing, even
if the condition $E_J \ll E_C$ is satisfied.

It is useful to discuss the relevant scale of energies and temperatures in a
specific example. Let us consider a system of $10^3$ atoms per well and the
value $\mu = 5 \hbar \omega_{ho}$ for the chemical potential. The coupling
constant $E_c$ is about $5 \times 10^{-3} \hbar \omega_{ho}$. By choosing $%
E_J=\hbar\omega_{ho}$ one has $\omega_p \sim 0.1 \omega_{ho}$. With the
above choices both the conditions $\omega_p \ll \omega_{ho}$ and $E_C \gg
E_J/N^2$ needed to apply the formalism developed above are well
satisfied. At $T=0$ the system is practically coherent ($\alpha \simeq 1$)
since $E_C \ll E_J$. However if $T$ is of the order of the oscillator energy 
$\hbar \omega_{ho}$, which corresponds to a rather low temperature in
standard experiments, the coherence is partially lost because of thermal
effects. By changing the value of $E_J$ and/or $T$ one can clearly explore
other interesting regimes.

The Josephson effect with two condensates can be naturally extended to an
array of multiple wells. This extension is particular relevant because of
the recent experimental efforts  to investigate 
one-dimensional optical lattices 
\cite{kas}. The array of condensates confined in multiple wells is described
by the multiple-condensate Hamiltonian 
\begin{equation}
\hat{H}_{J}=-{\frac{1}{4}}E_C\sum_{k=1}^{N_s}\frac{\partial ^{2}}{\partial
\Phi_k ^{2}}-E_{J}\sum_{k=1}^{N_s-1}\cos (\Phi_{k+1}-\Phi_k)  \label{HJarray}
\end{equation}
where $N_s$ is the number of wells, $\Phi_k$ is the phase of the $k$-th
condensate and $E_C$, $E_J$ are the parameters characterizing the
double-well Hamiltonian (\ref{HJquant}). In the large $N_s$ limit the
one-dimensional Hamiltonian (\ref{HJarray}) is known to exhibit a phase
transition at zero temperature occurring at the critical value $E_C = 1.62
E_J $ \cite{bradley}. The "superfluid" phase is not
accompanied by the appearence of an order parameter and cannot be described
in terms of mean field theories. It is instead characterized by an algebraic decay
of the phase correlation function $\langle \cos(\Phi_k-\Phi_l)\rangle$ at
large distances $\mid k-l\mid \gg 1$ (quasi long range order). The decay becomes
exponential for $E_C >1.62 E_J$. The phase transition disappears at finite
temperature. Actually in the "classical" regime of high temperatures the
phase correlation functions can be evaluated analytically and one finds the
simple result 
\begin{equation}
\langle \cos(\Phi_k-\Phi_l)\rangle_{cl} = \alpha_{cl}^{\mid k-l\mid}
\label{alphakl}
\end{equation}
which exhibits an exponential decay for large values of $\mid k-l\mid$. In eq.(\ref
{alphakl}) $\alpha_{cl}$ is the classical value (\ref{alphacl}) calculated
for the double well problem. Result (\ref{alphakl}) can be used to evaluate
the momentum distribution $n({\bf p})$ of the array. By taking the Fourier
transform $\Psi({\bf p}) = (2\pi\hbar)^{-3/2}\int d{\bf r} \Psi({\bf r}) 
\exp[-i{\bf p \cdot r}/\hbar]$ of the order parameter $\Psi({\bf r}) =
\sum_{k=1}^{N_s} \Psi_k({\bf r})e^{i\Phi_k}$ and using the relationship $%
\Psi_k({\bf p}) = \Psi_l({\bf p})\exp[-i(k-j)p_xd/\hbar]$ following from the
translational invariance properties of the array, one finds the result 
\begin{equation}
n({\bf p}) = n_0({\bf p})\sum_{k,l} e^{-i(k-l)p_xd/\hbar}\langle
\cos(\Phi_k-\Phi_l)\rangle
\end{equation}
\begin{equation}
= N_sn_0({\bf p}){\frac{1-\alpha_{cl}^2 }{1+\alpha_{cl}^2-2\alpha_{cl}%
\cos(p_xd/\hbar)}}  \label{npA1}
\end{equation}
where $n_0({\bf p})$ is the momentum distribution of each condensate, $d$ is
the distance between two consecutive condensates and, in deriving the last
identity, we have assummed $N_s \gg 1(1-\alpha_{cl})$. Result (\ref{npA1})
explicitly points out the effects of coherence. When $\cos(p_xd/\hbar)=1$
the incoherent signal $N_sn_0({\bf p})$ is amplified by the factor $%
(1+\alpha_{cl})/(1-\alpha_{cl})$. Viceversa, when $\cos(p_xd/\hbar)=-1$ the
signal is suppressed by the factor $(1-\alpha_{cl})/(1+\alpha_{cl})$.

In conclusion we have investigated the consequences of the quantum and thermal
fluctuations of the phase on the coherence phenomena exhibited by Josephson
like configurations. Thermal effects are predicted to play a major role even
at very low temperature and should be consequently taken into account in
order to control the transition to the new quantum phases exhibited by these
ultracold Bose gases.

\bigskip

Useful discussions with M. Inguscio, M.A. Kasevich and A. Smerzi are acknowledged.
This research is supported by the Ministero della Ricerca Scientifica e 
Tecnologica (MURST).

Fig.1 Coherence factor $\alpha$ at zero temperature 
as a function of the ratio $E_C/E_J$.

Fig.2 Coherence factor $\alpha$ 
as a function of the ratio $T/E_J$ for  $E_C/E_J =3$ (a), $E_C/E_J=1$ (b). 
The classical curve (\ref{alphacl}) is also shown (c).


\begin{references}

\bibitem{mitinterf}  M.R. Andrews, C.G. Townsend, H.-J. Miesner, D.S.
Durfee, D.M. Kurn, and W. Ketterle, Science {\bf 275}, 637 (1997).

\bibitem{leggett}  A.J. Leggett, Rev. Mod. Phys., April 2001.

\bibitem{jila} D.S. Hall, M.R. Matthews, C.E. Wieman, and E.A. Cornell,
Phys. Rev. Lett. {\bf 81}, 1542 (1998).

\bibitem{kas}  B.P.~Anderson and M.A.~Kasevich,  Science {\bf 282}, 1696
(1998); C.\ Orzel, A.K.~Tuchman, M.L.~Fensclau, M.~Yasuda, and
M.A.~Kasevich, Science {\bf 291}, 2386 (2001).

\bibitem{java}  J.\ Javanainen, M.\ Ivanov, Phys. Rev. A {\bf 60}, 2351
(1999).

\bibitem{smerzi}  A.~Smerzi, S.~Fantoni, S.~Giovannazzi, and S.R.~Shenoy,
Phys. Rev. Lett., {\bf 79}, 4950 (1997)

\bibitem{zapata}  I.\ Zapata, F.\ Sols, and A.J.\ Leggett, Phys. Rev. A, 
{\bf 57}, R28 (1998)

\bibitem{note1}  If the relative phase is localized around a value $\Phi
_{0}\ne 0$ one can simply use (\ref{URphi}) by replacing $\Phi $ with $\Phi
-\Phi _{0}$

\bibitem{PS}  L.P.\ Pitaevskii, S.\ Stringari, Phys. Rev. Lett. {\bf 83},
4237 (1999).

\bibitem{mitnp}  J.~Stenger, S.~Inouye, A.P.~Chikkatur, D.M.~Stamper-Kurn,
D.E.~Pritchard, and W. Ketterle, Phys. Rev. Lett. {\bf 82}, 4569 (1999).

\bibitem{RMP}  F. Dalfovo, S. Giorgini, L.P. Pitaevskii, and S. Stringari,
Rev. Mod. Phys. {\bf 71}, 463 (1999).

\bibitem{ADS}  J.R~Anglin, P.~Drummond, and A.~Smerzi, cond-mat/0011440


\bibitem{bradley}  R.M. Bradley and S. Doniach, Phys. Rev. B {\bf 30}, 1138
(1984)
\end{references}
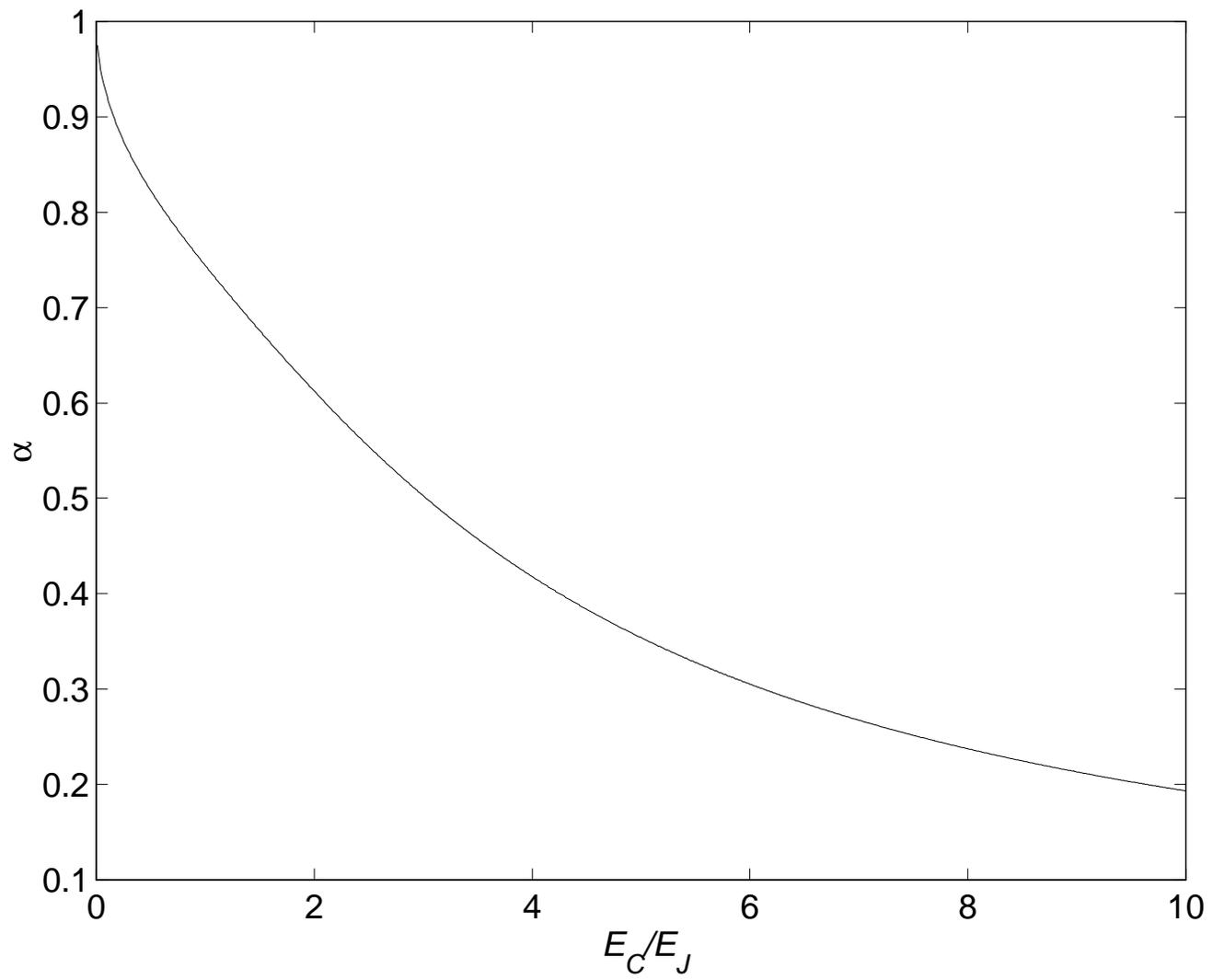
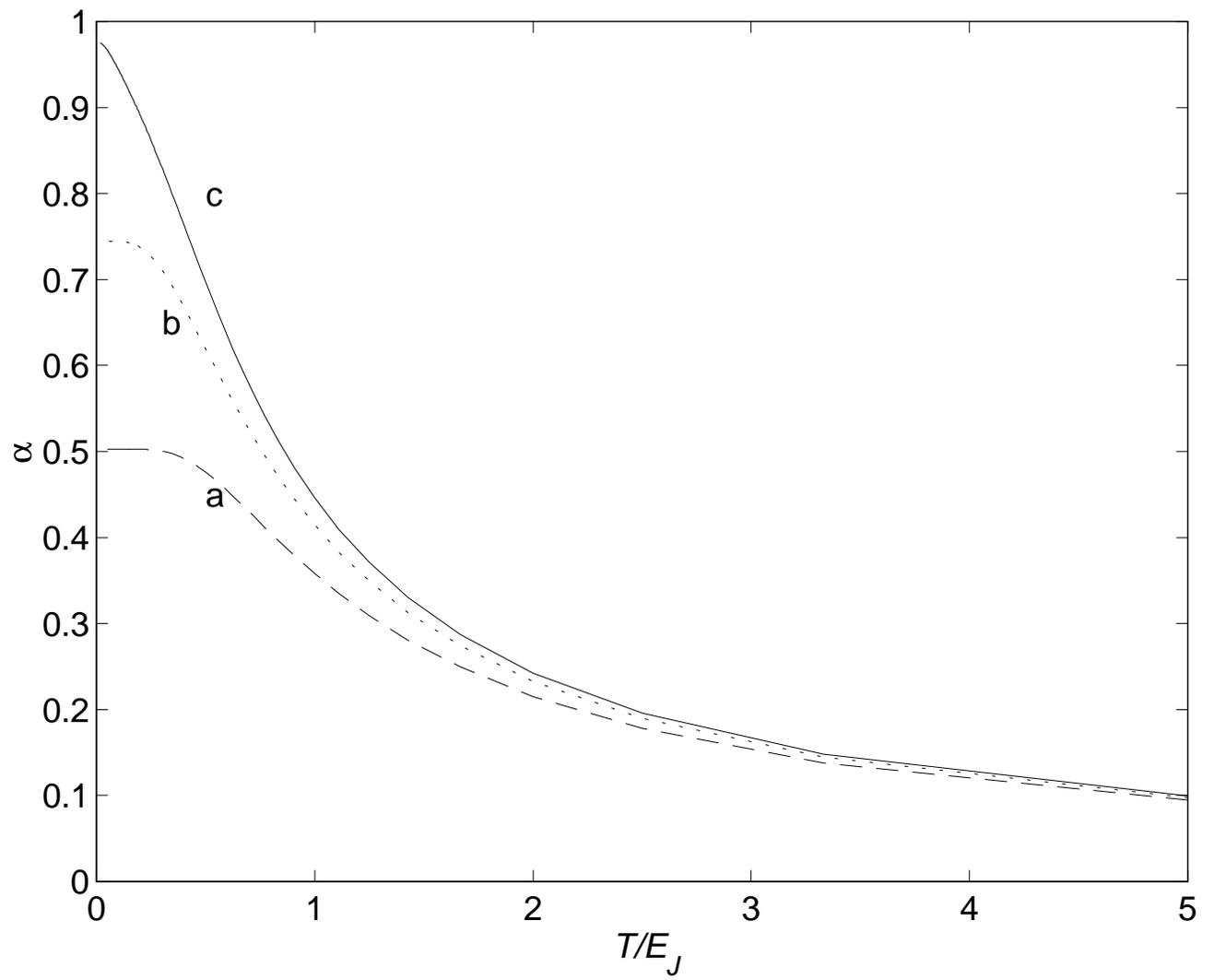
\end{document}